\documentclass[manuscript]{acmart}
\usepackage{tabularray, hhline, array, multirow}
\AtBeginDocument{%
  \providecommand\BibTeX{{%
    \normalfont B\kern-0.5em{\scshape i\kern-0.25em b}\kern-0.8em\TeX}}}

\setcopyright{acmcopyright}
\copyrightyear{2018}
\acmYear{2018}
\acmDOI{XXXXXXX.XXXXXXX}

\begin{document}
\title[Shaping Online Dialogue]{Shaping Online Dialogue: Examining How Community Rules Affect Discussion Structures on Reddit}

\author{Anna Fang}
\affiliation{%
 \institution{Carnegie Mellon University}
 \city{Pittsburgh}
 \state{Pennsylvania}
 \country{USA}}

\author{Wenjie Yang}
\affiliation{%
 \institution{Carnegie Mellon University}
 \city{Pittsburgh}
 \state{Pennsylvania}
 \country{USA}}

 \author{Haiyi Zhu}
\affiliation{%
 \institution{Carnegie Mellon University}
 \city{Pittsburgh}
 \state{Pennsylvania}
 \country{USA}}
 
\renewcommand{\shortauthors}{Fang et al.}

\begin{abstract}
Community rules play a key part in enabling or constraining the behaviors of members in online communities. However, little is unknown regarding whether and to what degree changing rules actually affects community dynamics. In this paper, we seek to understand how these behavior-governing rules shape the interactions between users, as well as the structure of their discussion. Using the top communities on Reddit (“subreddits”), we first contribute a taxonomy of behavior-based rule categories across Reddit. Then, we use a network analysis perspective to discover how changing implementation of different rule categories affects subreddits’ user interaction and discussion networks over a 1.5 year period. Our study find several significant effects, including greater clustering among users when subreddits increase rules focused on structural regulation and how restricting allowable content surprisingly leads to more interactions between users. Our findings contribute to research in proactive moderation through rule-setting, as well as lend valuable insights for online community designers and moderators to achieve desired community dynamics.
\end{abstract}



\maketitle

\section{Introduction}
Online communities have become increasingly integral in people’s everyday lives for building social relationships and discussing any topic of interest. However, an ongoing challenge in the creation and maintenance of these communities is moderating unwanted user behaviors, such as preventing online harassment or mitigating misinformation \cite{Budak2011-vj, Kowalski2015-zh}. Practices for achieving content moderation can include anything from manual human filtering of unwanted content \cite{Arsht2018-bk, Steiger2021-qh} to automated tools and bots \cite{Chandrasekharan2019-no, Gollatz2018-sq, Myers_West2018-oh, Seering2018-cz}. However, while most large platforms (e.g. Facebook, YouTube, Reddit) engage in multiple levels of content moderation, perhaps the most fundamental element of online community moderation is community rules. 

Community rules can encode platform and community values \cite{Fiesler2018-cy, Weld2021-eq}, form desired behaviors \cite{Fiesler2018-cy, Kraut2012-hv}, and act as both a guide and explanation for moderators’ decisions \cite{Seering2020-gp}. To quote Gillespie in his formative work on online content moderation: "\textit{There is no platform that does not impose rules, to some degree.}" \cite{Gillespie2018-ax}. Community governance and rules can take the form of anything from officially written rules to unspoken norms, and from overarching platform-level policies to customized sub-community rules \cite{Cialdini_undated-ed}. Past evidence shows that community-created rules based on existing norms are among the most effective for regulating users' behaviors, easiest for new members to adopt, and are context-sensitive to unique community goals and settings \cite{Fiesler2018-cy,Ostrom2000-uq,Seering2020-gp}. However, while explicit and community-created rules are effective for regulating user behaviors, there lacks empirical evidence regarding how and to what degree they affect user behavior, including changing the ways in which users interact or discuss topics. As a result, in this study, we seek to understand what behavior-governing rules are implemented across Reddit, one of the most popular social media platforms, and the effects of these rules on user interactions and discussion structures. 

Reddit, which largely relies on community-specific rules for its moderation, is the 7th most-visited website in the US and has over 130,000 active online sub-communities called “subreddits”. Since 2008, subreddits have been created and customized by Reddit users for nearly every topic of interest, from discussing technology (i.e. r/gadgets) and relationships (i.e. r/relationship\_advice), to users posting selfies of themselves to be insulted by other members (r/RoastMe). Reddit users primarily interact through creating their own posts, commenting, and upvoting or downvoting others' posts and comments. Users are largely anonymous on Reddit, with profiles containing only sparse information and rarely any profile photo. Although Reddit includes platform-wide policies (i.e. terms and conditions), its primary source of governance is each subreddit’s rules created by their own community moderators; these rules are formalized and visible most commonly in the subreddit’s sidebar. The rich and diverse community-driven rule system of Reddit has yielded a variety of online community research, including studies on collective behavior, social movements, moderation, and psychology of users \cite{Chandrasekharan2018-jh, De_Choudhury2014-yc, Jamnik2017-ee, Jhaver2019-yy, Medvedev2019-df}. Reddit's large user-base and diverse subcommunities makes the platform fitting for our work's analysis in comparing and contrasting the relationship between communities' rules and user behavior.

Our study is two-fold. First, we take inventory of what rules govern user behavior across Reddit through  a thematic analysis of Reddit’s top communities. Second, we analyze the most popular subreddits over a 1.5 year time period to investigate how changes in community rules impact discussion patterns through a network analysis approach, which has been used by researchers for decades to understand online community users' behaviors \cite{Ahuja1999-bj,Buntain2014-az,Datta2019-dc,Garton1997-fg,Kim2018-kf}. As opposed to focusing on language use as past work has done, we contribute to analyzing moderation effects from a social network analysis lens, which can be equally key for explaining behavior patterns (e.g. users’ involvement in the community, relationship structures) important for community designers and moderators \cite{Garton1997-fg}. 

In particular, we answer the following research questions:

\textbf{RQ1: What are the rules on Reddit that govern member behavior?}

\textbf{RQ2: How do changes in rules affect communities’ discussion and user interaction structures?}\\

Our study finds that behavior-governing rules on Reddit are covered by 15 different rule types and can be organized into three high-level rule themes: Community Values, Content Restrictions, and Structural Regulation. Additionally, we find that changes in these rule themes cause significant changes in user interaction and discussion structures. For example, we find that, surprisingly, greater emphasis on restricting allowable content  leads to higher levels of user interaction rather than limiting interaction or posting behaviors. We also found that increasing the structure of formatting and posting in a subreddit raises clustering of users as well as leads to more distributed interaction in the community. Our study provides a taxonomy for research to study the effects of Reddit rules on user behavior, as well as contributes to the rising interest in proactive moderation \cite{Schluger2022-il,Seering2020-gp} and encouraging behaviors through rule- and norm-setting \cite{Butler2008-aa,Chandrasekharan2018-jh,Fiesler2018-cy,Gillespie2018-ax,Ostrom2000-uq,Weld2021-eq}. Our findings also lend significant knowledge to community creators and designers of online communities on how particular types of rules can achieve desired community dynamics.

\section{Related Work}
In this section, we cover prior work done on the rules and moderation of online communities, as well as a review of how network analysis has been applied to the online community context. Our study bridges the gap in understanding how rule-setting affect online communities, and provide value in guiding the future of proactive moderation and network analysis perspective in content moderation research.
\vspace{-0.5em}
\subsection{Moderation of Online Communities}
With the rise of social media, moderation has also grown in importance as well as difficulty (Gillespie, 2018). Moderation efforts have targeted a number of diverse issues, including the spread of misinformation \cite{Caplan2018-iy, Sharevski2021-mg}, hateful speech and harassment \cite{Jhaver2018-ij, Jimenez_Duran2021-ie, Kwak2015-vr}, and spam \cite{Alberto2015-qs,Seering2017-ow}. Moderation can affect entire communities through processes like quarantining and banning \cite{Chandrasekharan2017-nn, Saleem2018-hl} as well as individuals, such as through deplatforming controversial users \cite{Jhaver2021-qh}. In recent years, there has also been a rising interest for using automated and artificial intelligence-backed methods to filter and adjust content \cite{Gillespie2020-au}. For example, HCI researchers have explored the opportunities for leveraging AI-based moderation in social VR \cite{Schulenberg2023-sf}, built recommendation systems for moderating online communities \cite{Chandrasekharan2019-no}, and alert bots for potentially violating content \cite{Chandrasekharan2019-no, Kiene2020-eq}. However, researchers have also explored the trade-offs and downfalls in using automatic forms of moderation, including challenges to human agency and balancing human-AI collaboration efforts \cite{Jhaver2019-yy, Jiang2019-md, Mackeprang2019-ir}. 

In terms of community rules, guidelines and policies are vital for reflecting community values into member behaviors. Although community rules have historically been set by broad platform-level policies (i.e. Terms of Service), these are often difficult to digest for members and may be largely disobeyed \cite{Fiesler2018-cy}. On the other hand, informal norms of communities play an important role in regulating behavior for both global and local community values but are difficult for new members to adopt and can result in high rates of drop-out if violated \cite{Chandrasekharan2018-jh, Lampe2014-zm, Seering2017-ow}. Norms are emergent and evolve over time in a community, and are often turned into explicit rules on platforms like Reddit that have local-level moderation \cite{Chandrasekharan2018-jh}. As a result, rules such as those on Reddit that reflect norms on a local level but are also formalized and visible in the community are an effective means of moderation and prevent unwanted behavior \cite{Lampe2014-zm, Matias2016-po, Mills2015-nj, Ostrom2000-uq}. However, subreddit rules can also cause tension within a community, such as rules that deter contribution to the community versus ensuring positive and high-quality content \cite{Kraut2012-hv, Lin2017-ff}.

We highlight especially relevant work done by Fiesler et al. that categorized the rules of Reddit. Fiesler et al. characterized the types and frequency of subreddit rules at scale, finding both context-dependent and sitewide commonalities across the platform \cite{Fiesler2018-cy}. However, we note key limitations that make this prior classification unfit for our study, including its broad rule categories and its categorization being primarily content-based rather than behavior-based. Some examples of this are that any rule mentioning promotional content (regardless of how the rule dictates user behavior with it) goes into an “Advertising \& Commercialization” category, and any rule mentioning requirements of a community would be categorized into a broad “Prescriptive” category. Additionally, two rules such as “be civil” versus “don’t be mean” can be categorized into different categories (“prescriptive” vs. “restrictive”) despite having essentially the same intent for a community’s users. As a result, our work builds upon Fiesler et al. by contributing a classification of Reddit’s rules through a behavior-based lens, which can be utilized for studies on the actual effects of rules on user behaviors and patterns.

\subsection{Network Science Applications to Online Communities}
While there has not been past work studying the effects of rules on community discussion through a network analysis lens, past work in network science and social networks has often centered on elements such as information diffusion and detection of influential actors \cite{Newman2004-nw,Newman2004-wl}. Past work has been able to identify certain leadership or social roles in online communities. Buntain and Golbeck demonstrated that a common social role, the “answer-person”, can be automatically identified on Reddit through their user interactions in the communities \cite{Buntain2014-az}. Similarly, Zhang studied the topology of an online help-seeking community through social network analysis methods, finding that a series of network-based ranking algorithms could identify users of high expertise in the community \cite{Zhang2007-ln}. Importantly, this identification of influential roles does not need to rely on expensive content analysis methods but rather can be performed with just the interaction structure of the network. Still, much work relies on computational social science methods to do this identification, using characteristics such as karma scores and users’ posting characteristics to determine opinion leadership \cite{Kilgo2016-wq}.

There has been some limited past work on exploring Reddit's network structure specifically. For example, Newell et al. conducted a study examining Reddit during a period of community unrest in 2015, finding that Reddit as a whole still persevered against large-scale changes in user behavior and possible migration of users to alternative platforms \cite{Newell2016-yl}. In terms of the discussion structure on Reddit, comment threads were found to resemble a topical hierarchy such that top-level comments begin a subtopic that then proceeds further down the comment levels. Because of this topical hierarchy, Reddit comments could be used for further implications such as web search \cite{Weninger2013-xd}. Political discussions on Reddit, which are often intensive and controversial, also provide a rich ground for identifying patterns of disputes and disruptions in discussion threads \cite{Guimaraes2019-hr}. Other work has focused on comparing network structures of related communities on Reddit, such as contrasting the benefits and challenges in different teaching-related subreddits \cite{Staudt_Willet2020-mw} or communication patterns in pro-eating disorder versus pro-recovery communities \cite{Fettach2020-fh}.

\section{Data}
In order to analyze how community rules affect online discussion, our study required two main data elements: (1) community rules and their changes over time, and (2) subreddits’ posts and comments. Below, we describe our data collection process using Reddit API, the internet archive Wayback Machine \footnote{https://archive.org/web/}, and existing Reddit data archives.

Altogether, given limitations on data sources as expanded upon below, this study is limited to a 1.5 year time period of April 1, 2018 to August 31, 2019. Additionally, given that the majority of Reddit's 2.8 million subreddits are too inactive to have a meaningful number of posts or subscribers for analysis, our study is limited to the top 2000 subreddits, whose names were collected using the Reddit API. These popular subreddits are mostly longer-standing communities that have sufficient user discussion taking place daily to offer substantive analysis for our study.

\subsection{Rules}
Although Reddit's official API provides current information for subreddits, it is not able to return historical data nor track changes over time. As a result, we used Wayback Machine, a digital internet archive tool, for accessing historical snapshots of subreddit homepages and extracting rule data. Wayback Machine is a public digital internet archive tool where users can manually request a “snapshot” of any page at any given time; note that, as a result, not every subreddit for our study has snapshots as availability is dependent on user(s) manual requests. We downloaded all available archives of the top 2000 subreddit's homepages from the Wayback Machine within April 1, 2018 and August 31, 2019. In total, 1033 subreddits out of the top 2000 subreddits did not have any snapshot data during our 1.5 year timeframe, leaving us with 967 subreddits available for analysis. After retrieving the snapshots, we applied regular expressions to parse rules from these archived pages. 

In total, we obtained 347,254 snapshots from 967 subreddits of the top 2000. Detailed statistics about our final dataset is presented in Table \ref{table:descr-stats-table}.

\begin{table}
\centering
\begin{tblr}{
  row{even} = {b},
  row{3} = {b},
  row{5} = {b},
  cell{1}{1} = {t},
  hline{1,6} = {-}{},
}
Number of subreddits        & 967\\
{Number of subreddits with any rule changes \\(April 2018 - August 2019)}    & 487 \par{}\\
Total number of snapshots   & 347,254     \\
Avg (mean) number of snapshots per subreddit  & 359\\
{Avg (median) number of snapshots per subreddit} & {59} 
\end{tblr}
\caption{Descriptive statistics for our final dataset.}
\label{table:descr-stats-table}
\end{table}

\subsection{Posts and Comments}
Given that the Reddit API does not supply past historical data as discussed above, we also sought an archive of subreddits' posts ("submissions") and comments to monitor discussion changes. To do this, we used the expansive Reddit archive dataset called PushShift \footnote{https://github.com/pushshift/google\_bigquery}, which is a public data archive that has logged Reddit submissions and comments in monthly data dumps since 2015 \cite{Baumgartner2020-od}. PushShift is currently updated up to August 2019. All PushShift Reddit data contains posts, comments, timestamp, and user ID info.\footnote{Note that, at the time of conducting our study, PushShift was publicly available data and updated to August 2019. However, we note that PushShift may no longer be updating due to ongoing changes by Reddit in 2023 to their data APIs. } We use this data to generate networks representing user interaction and discussion structures, as well as generate metrics on those networks, in order to measure any significant changes caused by rule changes (see Section 5).

\section{Study 1: What rules on Reddit govern member behavior?}
We first use thematic analysis methods to answer RQ1: What are the rules on Reddit that govern member behavior?

\subsection{Methods: Thematic Analysis}
To generate our codebook, we utilized the most recent rules of 100 of our dataset’s original 2000 subreddits (see Section 3 Data) to categorize Reddit rules into behavior-based categories and their high-level themes. 100 subreddits were chosen randomly and are provided in the appendix. Our categorization reached data saturation within 100 subreddits, such that more subreddits gave no new rule categories. 

For each subreddit, we used the Reddit API to scrape each subreddit’s community rules. Among the 100 chosen subreddits, there were 799 community rules total with individual subreddits ranging from having zero to fifteen community rules. We conducted a thematic analysis from Braun and Clarke \cite{Braun2012-sb} with an iterative analytic cycle consisting of: (1) coding all rules (2) amalgating codes (3) discussing codes as a team (4) highlighting high-level themes (5) writing and revising codebook memos. All rule categories are behavior-based such that each includes whitelisted or blacklisted behavior. Our analysis continued until saturation, during which prior to 100 subreddits new subreddits gave no new rule codes or themes. Our analysis ended with 56 codes, which were then organized into 15 axial-codes and summarized into 3 rule themes (Community Value, Content Restrictions, Structural Regulation). It is important to note that subreddits write their rules in very different ways - some subreddits “combine” regulations into a single community rule’s text while others may split up a rule into different parts. To account for this, in our codebook a listed rule can fulfill multiple codes. 

\begin{table}
\centering
\begin{tabular}{|>{\hspace{0pt}}m{0.135\linewidth}|>{\hspace{0pt}}m{0.227\linewidth}|>{\hspace{0pt}}m{0.455\linewidth}|>{\hspace{0pt}}m{0.09\linewidth}|} 
\hline
\multicolumn{1}{|>{\centering\hspace{0pt}}m{0.135\linewidth}|}{\textbf{Theme}} & \centering\textbf{Rule}    & \multicolumn{1}{>{\centering\hspace{0pt}}m{0.455\linewidth}|}{\textbf{Example}}        & \multicolumn{1}{>{\centering\arraybackslash\hspace{0pt}}m{0.09\linewidth}|}{\textbf{\# subreddits}}  \\ 
\hline
\multirow{5}{\linewidth}{\hspace{0pt}Community Values}  & Discourages divisiveness  & \textit{No excessive arguing about social or political topics (r/trashy)}\par{}\textit{No controversial posts (r/KamikazeByWords)}  & 22     \\ 
\cline{2-4}
   & Enforces respect for others        & \textit{No hate speech (r/gifs)}\par{}\textit{Be nice (r/tipofmyjoystick)}& 87     \\ 
\cline{2-4}
   & Avoid distressing material& \textit{No upsetting content (r/Portland)}\par{}\textit{No videos involving death (r/youtubehaiku)}      & 40     \\ 
\cline{2-4}
   & No promotion of bad behavior       & \textit{Comments promoting piracy with any linkes to pirate sites are banned (r/visualnovels)}  & 18     \\ 
\cline{2-4}
   & Enable peer feedback      & \textit{Report content via modmail or Reddit's reporting system (r/WWE)}  & 10     \\ 
\hhline{|====|}
\multirow{7}{\linewidth}{\hspace{0pt}Content Restrictions}       & Posts must be verifiable  & \textit{No unsubstantiated rumors (r/disney)}\par{}\textit{No pseudoscience or misinformation (r/ADHD)}  & 23     \\ 
\cline{2-4}
   & No jokes& \textit{No memes (r/sushi)}\par{}\textit{No jokes, memes, or cartoons (r/thalassophobia)}       & 26     \\ 
\cline{2-4}
   & Content must be original  & \textit{Duplicate or similar post will be removed (r/canada)}    & 41     \\ 
\cline{2-4}
   & No promotional content    & \textit{No referral links/begging/faucets/casinos/pumps (r/CryptoCurrencies)}\par{}\textit{No begging, trading, or selling (r/pcmasterrace)} & 52     \\ 
\cline{2-4}
   & Allow promotional content, with approval    & \textit{Keep promotion reasonable (r/seaofthieves)}\par{}\textit{Follow guidelines about self-promotion (r/Homebrewing)}   & 66     \\ 
\cline{2-4}
   & Posts must be on-topic    & Must be relevant to Overwatch (r/Overwatch)    & 79     \\ 
\cline{2-4}
   & Posts must be high-quality& New accounts with no karma will be removed to weed out spam (r/beauty)\par{}Posts or questions lacking substance may be removed (r/coffee)   & 5      \\ 
\hhline{|====|}
\multirow{3}{\linewidth}{\hspace{0pt}Structural Regulation}      & Requires a minimum text   & Posts must be at least 300 characters (r/ADHD) & 23     \\ 
\cline{2-4}
   & Requires template for posts        & Playlists can only be posted on weekends (r/LofiHipHop)\par{}Follow post template (r/MagicArena)& 33     \\ 
\cline{2-4}
   & Limits use of title versus description text & No description use (r/ar15)  & 2      \\
\hline
\end{tabular}
\caption{Rule codebook organized by theme,  including examples for each axial-code and the number of classified subreddits in the thematic analysis. }
\label{table:rule-codebook}
\end{table}
\vspace{-1em}
\subsection{Results: Rule Codebook}
Our codebook’s 15 axial-codes organized into 3 high-level rule themes are shown in Table \ref{table:rule-codebook}. The 3 themes are Community Values (member interactions and considerations), Content Restrictions (limits to post content), and Structural Regulation (requirements to a post’s structure and form). Below, we review each code and its definition.

\begin{enumerate}
\item \textbf{Community Values.} Rules under Community Values articulate basic values and considerations for other members in a community and generally encourage moral behavior. We provide sub-themes regarding rules relevant to Community Values described below:
\begin{enumerate}
\item Discourages divisiveness. Disallow divisive content and thus promote a community where members are harmonious and in agreement on the community’s topic.
\item Enforces respect for others. Encourage general respect for other members, including policies like ”No hate speech” or ”No personal insults”.
\item Avoid distressing material. Disallow posting generally unwanted material to members, such as NSFW (Not Safe For Work) or NSFL (Not Safe For Life) material.
\item No promotion of bad behavior. Disallow users to promote bad behavior, such as anything breaking the law.
\item Enable peer feedback. Allow for peer regulation such that general reporting tools are available to all members where they can input reports on other users and posts.
\end{enumerate}

\item \textbf{Content Restrictions.} Rules under Content Restrictions limit what is allowed in a post or comment by users.
\begin{enumerate}
\item Posts must be verifiable information. Enforces verifiable, factual, and legitimate information. Often, these subreddits remove posts if they are not based on reliable sources and often are news- or medical-related subreddits.
\item No jokes. Posts must not be satire, jokes, or memes.
\item Content must be original. Users are encouraged to post original content made by themselves, or check the subreddit to see if their post has been made previously.
\item No promotional content. Some subreddits disallow promotionals completely. Promotionals are usually defined as self-promotion (e.g. asking members to follow your social media account) or advertising of products and services.
\item Promotional content allowed with approval. Promotionals are allowed, but with limited approval through moderator team or under certain requirements.
\item Posts must be on-topic. Posts must be on-topic and in the spirit of the subreddit. For example, subreddit r/aww posts are not allowed to contain morbid or sad content due to the community’s purpose for sharing cute photos.
\item Posts must be high-quality. ”High-quality” posts include questions of substance, encourage discussion, are not spam, and/or have high-quality media. Some subreddits ensure posts are high-quality by explicitly banning new accounts to avoid spam content.
\end{enumerate}

\item \textbf{Structural Regulation.} Structural Regulation rules define the physical structure of a post, such as where the post is allowed to be in the community or a post’s formatting.
\begin{enumerate}
\item Requires a minimum text. Some subreddits encourage descriptive/long-form posts. A few subreddits even require a character- or word-count requirement.
\item Requires template for posts. Regulate the structure or timing of posting. Some communities restrict the location of a post, such as limiting posts on certain topics to be comments on designated ”threads”, while others limit the timing of certain types of posts to designated times and days of the week.
\item Limits the usage of title versus description text. Limit the usage of the title versus description text; for example, r/askReddit posts are limited to using the title area to ask a question and not the post’s body description.
\end{enumerate}
\end{enumerate}

\section{Study 2: How do changes in rules affect discussion?}
Based on our created codebook and its three themes that encompass Reddit’s rules, we then hypothesized and examined the relationship between changes in these rule themes and changes in communities’ discussion structures. Below, we answer our study’s RQ2: How do changes in rules affect communities’ discussion and user interaction structures?

\subsection{Hypotheses}
We formed hypotheses H1-H5 described below based on prior work.

We first hypothesize regarding how changes in community rules affect the total volume of user interaction in a given community. There is evidence that a positive, healthy, and supportive community environment based in mutual respect may benefit from a higher volume of interaction \cite{Ren2007-av}. Additionally, the rule theme of Community Values includes items like disallowing divisiveness or controversial topics; members are more likely to contribute to a community when high agreeableness and similarities exist \cite{Ling2005-xi}. Thus, we expect that adding rules regarding Community Values will result in greater volume of user interaction. It also follows intuitively that barriers to posting content would deter members from engaging in a community \cite{Ling2005-xi} such as restrictions on permissible content or additional effort needed to follow posting and formatting guidelines. As a result, we expect that subreddits adding rules in Content Restrictions or Structural Regulation will have lower levels of user interaction. Thus:
\\

\noindent\textbf{H1. Volume of User Interaction.} \textit{Communities with greater Community Values will have higher levels of user interaction, while communities with greater Content Restrictions or Structural Regulation will have lower levels of user interaction.}
\\

In addition to the total volume of user interaction, we also comment on the distribution of user interaction among a community. In addition to group-level dynamics being enhanced by positive interactions, leadership is also more likely to be shared given open and encouraged participation in a community (such as through prohibiting bad or disrespectful behaviors) \cite{Johnson2015-uz}. As a result, we hypothesize that greater emphasis on Community Values would lead to more evenly distributed interactions throughout the community as a whole, as well as greater “shared leadership” as more evenly distributed power. Additionally, it follows that increased Structural Regulation would result in smaller sub-groups, resulting in more local “leadership” – people who post or comment a lot in the community – across the community as a whole. Additionally, smaller online communities tend to have less dyadic interactions and more group-level dynamics involving more users contributing and participating \cite{Hwang2021-tp}. As a result, we hypothesize that greater Structural Regulation leads to smaller sub-groups, which leads to a more even distribution of interaction and power across the community.  
\\

\noindent\textbf{H2a. Distribution: User Interaction.} \textit{Communities with greater Community Values or Structural Regulation will have more even distribution of user interaction throughout the community.}\\
\textbf{H2b. Distribution: User Power.} \textit{Communities with greater Community Values or Structural Regulation will have more even distribution of power among users in the community.}
\\

We also analyze the relationship between community rules and overall connectivity measures. We hypothesize that requirements for posts to follow a certain post template or structure, such as directing posts to designated threads, would allow members to navigate the community and more easily find their topics of interest (and thus, other users also interested in similar topics).  As a result, a subreddit may naturally become split up into sub-groups (“sub-subreddits”) as users direct their interactions towards only a subset of the total community. As a result, we hypothesize that greater Structural Regulation leads to higher clustering, but overall lower density as users interact with a smaller proportion of the community. Additionally, rules that lead towards a more community-like structure, such as preventing divisive or malicious interactions (Community Values), may lead to less clustering of users (Newman and Park 2003). Similarly, communities with unbiased and higher quality content (Content Restrictions) may also see less divisiveness and subgroup formation. Therefore, we hypothesize:
\\

\noindent\textbf{H3a. Connectivity: Clustering.} \textit{Communities with greater Community Values or Content Restrictions will have overall lower clustering of users. However, communities with greater Structural Regulation will have higher clustering of users.}\\
\textbf{H3b. Connectivity: Density.} \textit{Communities with greater Structural Regulation will have lower density of users. }
\\

Although prior work on studying commitment and contribution to online communities has mostly focused on member retention \cite{Arguello2006-rc, Kraut2011-pl, Lin2017-ff}, there is also some literature measuring community success through content production \cite{Agichtein2008-ju, Dong2020-za, Tang2012-eo}. As a result, we also conducted analysis related to the volume and distribution of posting and commenting activity. Similar to hypothesis H1, greater restrictions on posting whether on content (Content Restrictions) or structure (Structural Regulation) may deter contributions to a community. Additionally, the Content Restrictions rule theme includes banning of spam and promotional content. Deterrence of spam content would intuitively lead to less posting or commenting volume. In terms of rules establishing healthy or agreeable interactions (Community Values), prior work on Wikipedia has found that contributing content is primarily motivated by expectations of reciprocity and reciprocity sustained continued interactions \cite{Xu2015-tf}. It may follow then that positive reciprocal behavior, which can be achieved through norms like avoiding distressing other users, kindness and respect, and agreement among members, would contribute to greater content generation. Additionally, there is evidence that values like community generousness, members having high self-esteem, and open expression are more likely to lead to content contribution \cite{Wang2003-ae}. Thus:
\\

\noindent\textbf{H4. Volume of posting activity.} \textit{Communities with greater Content Restrictions and Structural Regulation will have lower volume of posting activity, while communities with greater Community Values will result in higher volume of posting activity.}
\\

People engage more when content contains users debating or disagreeing with each other. For example, visible online conflict increases visual attention to the post \cite{Dutceac_Segesten2022-hf}. Disagreement may also lead to more engagement with a post as users contribute and continue to back-up their stance or otherwise engage with others’ opinions \cite{Qiu2015-ng}. It then follows that, compared to communities allowing or promoting debate among members, communities that deter any disagreement or controversial speech would have less posts and comment threads with particularly high volume. As a result, we hypothesize that communities with greater Community Values (which includes rules like disallowing hateful speech or disagreement among members) have a more even distribution of comments as a whole. 

In a similar vein, the absence of spam or trolling on particular posts may get rid of especially long comments and posting threads. As a result, greater Content Restrictions (including rules such as banning spam, promotional content, and low-quality content) would lead to more even distribution of comments as a whole. Thus:
\\

\noindent\textbf{H5. Distribution of commenting activity.} \textit{Communities with greater Community Values and Content Restrictions will have more even distribution of commenting activity across the community. }

\subsection{Methods}
Given H1-H5, we first extracted rules to analyze their changes (Section 5.2.1) based on our work from RQ1. Then, we formed networks for all subreddits and calculate their metrics (Section 5.1.2). Lastly, we present our causal inference and regression methods (Section 5.1.3)

\subsubsection{Automatic Rule Extraction and Change Detection}
\begin{table}
\centering
\begin{tabular}{lll} 
\hline
Model      & Accuracy & F1    \\ 
\hline\hline
Logistic Regression & 0.76     & 0.77  \\
Random Forest       & 0.65     & 0.71  \\
Decision Tree       & 0.67     & 0.67  \\
SVM        & 0.68     & 0.72  \\
XGBoost    & 0.73     & 0.76  \\
\hline
\end{tabular}
\caption{Accuracy and F1-score of models for our rule classifier. Our study used the highest accuracy Logistic Regression classifier for analysis.}
\label{table:model-acc}
\end{table}

After forming our hypotheses, we extracted rules from our dataset’s subreddits and identified changes over time. As we would not be able to manually code the vast amount of rules obtained from Wayback Machine (Section 3.1), we automated the coding process by transforming into a three-class classification task and implementing a machine-learning model to categorize the rules. We first annotated 701 individual Reddit rules to create our training set. This set consisted of 397 rules related to content restrictions (57\%), 241 centered on community values (34\%), and 63 about structural regulation (9\%). Subsequent preprocessing was carried out to clean the dataset, which involved removing punctuation, special characters, and stop words. These preprocessed rules were then converted into a 5000-dimensional vector representation via a bag-of-words model. Afterwards, we built a Logistic Regression classifier on 80\% of the training set and performed 10-fold cross-validation on the remaining test set. The model achieves average F1 scores of 0.78, 0.83, and 0.66 for types of community values, content restrictions, and structural regulation, respectively. The overall F1-scores of our final Logistic Regression classifier, as well as other attempted models, are displayed in Table \ref{table:model-acc}. By applying the classifier to all data, we discerned rule types and their counts at different time points for those subreddits that have archival records on the Wayback Machine.

For rule change detection, we scrutinized consecutive snapshots of each subreddit from the Wayback Machine to identify any shifts in the quantity of each rule type. This includes rule additions or removals within the three categories. Note that, as Wayback Machine archives web pages based on user requests instead of executing routine crawls, the intervals between snapshots can be inconsistent; as a result, change detection between two snapshots does not provide the precise timing of when the change occurred. To mitigate any substantial gaps between snapshots, we excluded change points that have gaps exceeding a certain length (i.e., 60 days).

Overall, we identified 948 rule change occurrences across 487 of the 967 subreddits with snapshots. The remaining 480 subreddits had snapshot data, but had stagnant rules during our study's 1.5 time period. More comprehensive statistics of these change points can be found in Table \ref{table:change-event}.

\begin{table}
\centering

\begin{tabular}{>{\hspace{0pt}}m{0.251\linewidth}>{\hspace{0pt}}m{0.225\linewidth}>{\hspace{0pt}}m{0.158\linewidth}} 
\hline
& \textbf{Number (Percentage)~} & \textbf{Mean (std)}  \\ 
\hhline{~==}
Change Event     & 948~  &      \\
\textbf{+ community values}  & 323 (34.07\%) & 1.41 (0.90) \\
\textbf{- community values}  & 153 (16.14\%) & -1.50 (1.04)\\ 
\hline
\textbf{+ content restriction}   & 527 (55.59\%) & 1.65 (1.20) \\
\textbf{- content restriction}   & 180 (18.99\%) & -1.44 (1.00)\\ 
\hline
\textbf{+ structural regulation} & 197 (20.78\%) & 1.24 (0.56) \\
\textbf{- structural regulation} & 88 (9.28\%)   & -1.17 (0.53)\\
\hline
\end{tabular}
\caption{Descriptive statistics about changes to rule types in dataset, including the number of change occurrences for each type and the average amount of change (e.g. +2 indicates adding two rules in the category, -1 indicates subtracted one rule in the category.}
\label{table:change-event}
\end{table}

\begin{table}
\end{table}

\subsubsection{Network Formation and Metrics}
Following identifying when rule changes occurred, we formed two kinds of networks to see how subreddits changed from before to after a rule change. Using posts and comments data from PushShift (Section 3.2), we created a user interaction network (representing how users interact with one another by commenting on each others' posts/comments) and a submission network (representing posting ("submitting") and each post's commenting activity) in order to test the above hypotheses in Section 5.1. We then used a matching-based method to find a control sample for each of the networks, and conducted regression analysis to find any significant relationships between types of rule changes and network metrics.

Below, we describe the construction of these networks as well as the metrics we calculated for these networks. Our network metrics were chosen based on past literature on online community metrics \cite{Borgatti2005-gt, Gruzd2013-dx, Kumar2006-ij,Pfeil2009-he} and potential interest to community moderators for relevance to their community.
\\

\noindent\textbf{User Interaction Network.} To form the user interaction network, a network G = \{V,E\} consists of nodes V and links E, where \textbf{nodes V are users in the subreddit} and the \textbf{directed edges of E are directed from a post's author to the recipient}. The user interaction network is a directed network representing user-to-user interactions. Each node in the network is a user in the subreddit and edges are an interaction between users; in the case of Reddit and this study, an interaction is a comment on another user's post or comment. Note that users on Reddit can also like or dislike other users' posts and comments, but this action is anonymous; there is no identification for who liked or disliked a post or comment and as a result likes/dislikes are not analyzable for this study.
We calculated the following network metrics for the user interaction networks:

\begin{enumerate}
    \item \textbf{Volume of User Interaction}: To measure volume of interaction, we consider the following four network metrics of vertex count, edge count, average in-degree centrality, and average out-degree centrality. 
        \begin{enumerate}
           \item \textbf{Vertex Count}: the number of vertices / nodes represents the number of active users in the subreddit.
           \item \textbf{Edge Count}: the number of edges between all vertices represents the total number of interactions occurring in the subreddit.
           \item \textbf{Average in-degree centrality}: For a given user X, the in-degree is the count of users who interacted (i.e. commented on someone’s post or comment) with X. We average the in-degree centrality amongst all nodes in the network.
           \item \textbf{Average out-degree centrality}: For a given user X, the out-degree is the count of users who X interacted with. We average the out-degree centrality amongst all nodes in the network.
        \end{enumerate}

    \item \textbf{Distribution of User Interaction}: In addition to measuring the volume of user interaction, we use the Gini coefficient to measure the equality or inequality of interaction throughout communities.
        \begin{enumerate}
   \item \textbf{Gini coefficient of degree centrality}: how equally interaction is distributed among users in the community.
   \item \textbf{Gini coefficient of betweenness centrality}: how equally power (betweenness centrality) is distributed among users in the community.
        \end{enumerate}

    \item \textbf{Connectivity}: We use density and clustering as proxies to measure how close and connected the users are in a subreddit, and how this connectivity may change as a result of rule alterations.
       \begin{enumerate}
  \item \textbf{Density}: Density is the proportion of connections existing relative to the total number of possible connections. High density may represent many connections among individuals
  \item \textbf{(Global) Clustering coefficient}: The global clustering coefficient is a metric of the degree to which community members tend to cluster together; specifically, global clustering coefficient (also known as transitivity) is a measurement of closed triplets over the total number of triplets (open and closed) in the network.
       \end{enumerate} 
\end{enumerate}
\vspace{\baselineskip} 
\noindent\textbf{Submission Network.}
The network G = \{V, E\} consists of nodes V and links E where \textbf{nodes V are posts (whether original submissions or a comment)} in the subreddit and the \textbf{unweighted, directed edges of E} are directed from a post to its parent post. The submission network is an unweighted and directed network representing the commenting and posting activity in a community. 

We calculated the following network metrics for the submission networks:

\begin{enumerate}
\item \textbf{Posting Activity}:
    \begin{enumerate}
        \item \textbf{Vertex Count}: the number of vertices / nodes represents the number of posts (“submissions”).
        \item \textbf{Edge Count}: the number of edges between all vertices represents the total commenting activity. Note that, in the submission network, nodes can have a maximum of one out-degree (a comment can only be on a single post/comment).
        \item \textbf{Average in-degree}: For a given submission X, the average in-degree is the average comments per post.
        \item \textbf{Gini coefficient of degree centrality}: A measure of equality or inequality of comments distributed throughout submissions in the community.
    \end{enumerate}

\end{enumerate}

\subsubsection{Control Subreddits}
Given that just using regression analysis cannot draw causal claims, we implemented a matching-based process inspired by prior work done in propensity score matching and case-control matching \cite{Caliendo2008-tz, Pearce2016-ss}. Using a similar approach as prior work done on Reddit specifically \cite{Chandrasekharan2022-mx}, we conducted a quasi-experimental analysis to identify causality between rules and network outcomes by using a matching-based process to choose a “control” for each “treatment” point – i.e. a subreddit that had a rule change. In order to account for individual differences between subreddits, we match each observation in the treatment group with an observation in the control group based on similarity of confounding characteristics, where the treated observation had a rule change while the control did not. 

In particular, we used the 487 subreddits that had any snapshot and rule change occurrences as a treated group. Then, for each treated subreddit, we selected its control as the subreddit that did not have any rule change during the treated subreddit’s observation period and had the most \textit{co-posting} of users – that is, the subreddit that had the most number of active users who posted in both the control and treated subreddit. We used co-posting based on prior work \cite{Chandrasekharan2022-mx} to control for important confounding characteristics like topic. After choosing control subreddits, our team did a manual review of 10 subreddits to verify that treatment and control subreddits had similar topics.

\subsubsection{Regression}
After the matching process, we used fixed-effects regression on the treatment-control pairs. We used fixed-effect regression to identify how changes in rules over time relate to changes in the network metrics described above. Our data is panel data, where each row represents a possible change point in community rules along with the type of change (e.g. Structural Regulation rules reduced by 1, Community Values rules gained by 1). In our fixed-effects regression, each group was set to be the treatment subreddit and its respective control subreddit. Our regression's dependent variable is the change in each network metric (i.e. the pre-rule change network metric subtracted from the post-rule change network metric), while the independent variables are changes in rule implementation (i.e. +1 Structural Regulation, +0 Community Values, -1 Content Restrictions). We also added in the number of subscribers in each subreddit as a control variable in the regression.

\subsection{Results: Regression Analysis}

\begin{table}
\renewcommand{\thetable}{\arabic{table}a}
\centering
\def\arraystretch{1.2}

\begin{tabular}{>{\hspace{0pt}}m{0.163\linewidth}>{\hspace{0pt}}m{0.221\linewidth}>{\hspace{0pt}}m{0.221\linewidth}>{\hspace{0pt}}m{0.221\linewidth}} 

      & \multicolumn{3}{>{\centering\arraybackslash\hspace{0pt}}m{0.663\linewidth}}{\textbf{Volume of User Interaction }\par{}H1}\\ 
\cline{2-4}
      & \textbf{Model 1} \par{} vertex count \par{}(log) & \textbf{Model 2} \par{} edge count \par{} (log) & \textbf{Model 3} \par{} degree, avg.\par{}  \\ 
\cline{2-4}
      & $\beta$ (SE)  & $\beta$ (SE) & $\beta$ 
      (SE)      \\ 
\hhline{~===}
\textbf{comm.~values}   & 0.028 (0.056)& 0.059 (0.065)        & \textbf{0.799*} (0.401)     \\
\textbf{content~restr.} & 0.025 (0.042)& 0.030 (0.048)        & \textbf{0.573*} (0.291)     \\
\textbf{struct.~reg.}   & 0.092 (0.087)& 0.149 (0.097)        & 0.318 (0.656)    \\ 
\hline
\textit{* p $<$ 0.05}   &     &    &\\
      &     &    &        
\end{tabular}
\label{table:user-reg-analysis-a}
\caption{Fixed-effect regression analysis results for user interaction networks (continued). Independent variables are rule types (community values, content restrictions, and structural regulation) while Dependent variables are the respective network metrics.}
\end{table}

\begin{table}
\renewcommand{\thetable}{\arabic{table}b}
\ContinuedFloat
\centering
\def\arraystretch{1.2}

\begin{tabular}{>{\hspace{0pt}}m{0.133\linewidth}>{\hspace{0pt}}m{0.147\linewidth}>{\hspace{0pt}}m{0.147\linewidth}>{\hspace{0pt}}m{0.001\linewidth}>{\hspace{0pt}}m{0.147\linewidth}>{\hspace{0pt}}m{0.147\linewidth}}
      & \multicolumn{2}{>{\centering\hspace{0pt}}m{0.374\linewidth}}{\textbf{Distribution of User Interaction }\par{}H2a, H2b} &  & \multicolumn{2}{>{\centering\arraybackslash\hspace{0pt}}m{0.374\linewidth}}{\textbf{Connectivity }\par{}H3a, H3b}  \\ 
\cline{2-3}\cline{5-6}
      & \textbf{Model 4}\par{}degree, \par{}gini. & \textbf{Model 5}\par{}betw. centr., \par{}gini.       &  & \textbf{Model 6}\par{}density\par{} & \textbf{Model 7}\par{}clustering\par{}~        \\ 
\cline{2-3}\cline{5-6}
      & $\beta$ (SE)   & $\beta$ (SE)  &  & $\beta$ (SE)      & $\beta$ (SE)    \\ 
\hhline{~==~==}
\textbf{comm.~val.}     & -0.001 (0.001) & 0.000 (0.000) &  & -0.039 (0.043)    & 0.001 (0.001)   \\
\textbf{content~restr.} & 0.001 (0.001)  & 0.000 (0.000) &  & -0.002 (0.032)    & -0.000 (0.001)  \\
\textbf{struct.~reg.}   & \textbf{-0.010*~}(0.002)& 0.001 (0.001) &  & -0.044 (0.072)    & \textbf{0.048*~}(0.002)      \\ 
\hline
\textit{* p $<$ 0.05}   &       &      &  & &       
\end{tabular}
\caption{Fixed-effect regression analysis results for user interaction networks (continued).}
\label{table:user-reg-analysis-b}
\end{table}

Our results from regression are reviewed below organized by hypothesis. In addition to analyzing types of rule changes, we also analyzed effects of changing the raw number of rules. We found that a greater number of rules overall led to more clustering and user interaction; full results are available in our appendix. We used statistical software package Stata for all regression analysis.

\subsubsection{Volume of User Interaction}

First, we review results for \textit{H1: Communities with greater Community Values will have higher levels of user interaction, while communities with greater Content Restrictions and Structural Regulation will have lower levels of user interaction.}

As displayed in Table 5, we found evidence to partially support H1 in that communities with greater Community Values have a significantly higher level of user interaction. However, surprisingly, we did not see support for H1 communities in terms of Content Restrictions and Structural Regulation; instead, we see the opposite expected effect as greater Content Restrictions leads to higher levels of user interaction. There were no significant effects seen for Structural Regulation. 

A rise in Community Values rules leading to more user interaction follows prior work such as Ren et al. that finds shared identity such as through disclosure and liking for one another \cite{Cartwright_undated-wm, Ren2007-av}can provide many opportunities for more social interaction and greater willingness for contribution. Additionally, there is reason to believe that this effect is cyclical – users who like each other are more likely to interact more, and interacting more leads to closer social connections \cite{Ren2007-av, Utz2003-up}

\subsubsection{Distribution of User Interaction}
Our results for {H2a: Communities with greater Community Values or Structural Regulation will have more even distribution of user interaction throughout the community} showed no significant results for Community Values, but did show support for Structural Regulation. In particular, an increase of one rule in Structural Regulation led to an almost 1\% more even distribution (i.e. a lower gini coefficient of degree centrality). 

Although we saw that distribution of user interaction is affected by rule change (Structural Regulation), distribution of power did not show any significant results. We saw no support for \textit{H2b: Communities with greater Community Values or Structural Regulation will have more even distribution of power among users in the community.}.

\subsubsection{Connectivity}
In terms of connectivity, we answered whether \textit{H3a. Communities with greater Community Values or Content Restrictions will have overall lower clustering of users. However, communities with greater Structural Regulation will have higher clustering of users.} and \textit{H3b: Communities with greater Structural Regulation will have lower density of users. }

H3a was only partially supported and H3b was not supported. A rise in Structural Regulation led to an almost 5\% increase in clustering. Only Structural Regulation among the 3 rule themes has a significant relationship with clustering as no significant effects were seen for Community Values and Content Restrictions. However, although Structural Regulation led to greater clustering, it did not lead to higher density.

We reason that subreddits that enforce structure in their community enable users to find content they are interested in and may create local groups/clusters around that topic of interest as a result; in other words, organized structure makes it easier for users to interact more with a clique of similar users. In particular, using a sample of 100 subreddits and running an OLS regression on binary independent variables for implementation of the three axial-codes (see Section X) under Structural Regulation (i.e. 1 if a subreddit implemented the code and 0 otherwise), we found that the only statistically significant relationship was seen for the code ”Require template for posts”. This axial code is fulfilled by a subreddit if it has rules covering one or more of (1) requiring all posts to follow a specific template, (2) requiring all posts to have a flair (i.e. a category) that users can filter by, and (3) directing posts to either be put in specified threads or posted during a specific time frame. A qualitative look at the subreddits that implement the code ”Require template for posts'' revealed that most of these communities revolved around broad subjects (e.g. r/portland) that may benefit from directing certain types of posts to certain threads or time periods (i.e. dedicated threads to buying/selling items).

The insignificant results for Community Values may stem from the possibility that communities without community-oriented rules already have a community-oriented culture of mutual respect and agreeability. Communities that do not implement rules disallowing divisive content or distressing material may not suffer from this content in the first place. Given that subreddits are inherently ”echo chamber-like” given that Reddit users self-select which communities to join in the first place, it is possible that all members in some subreddits are already in agreement with one another on topic and community goals.

\subsubsection{Volume of Posting Activity}
We next review results using submission networks. 

Our results for \textit{H4. Communities with greater Content Restrictions and Structural Regulation will have lower volume of posting activity, while communities with greater Community Values will result in higher volume of posting activity.} show no significant effects for Content Restrictions or Community Values. However, Structural Regulation led to significantly greater posting activity.

There may also be effects that cancel each other out. For example, Ren et al. \cite{Ren2007-av} has suggested that content restriction can provide opportunity for greater self-disclosure, such as people talking about themselves or their day offhand, and closer friendships. However, this may be intimidating or confusing for newcomers, and even violate their expectations of the community – thus decreasing their overall involvement in the community. 

\subsubsection{Distribution of Posting Activity} Lastly, we answer
\textit{H5: Communities with greater Community Values and Content Restrictions will have more even distribution of commenting activity across the community. }

Although we see partial support for hypothesis H5 as commenting activity becomes more evenly distributed with a rise in Content Restrictions, Community Values had no significant effect. However, a rise of Content Restrictions predicts a small (~.2\%) more even distribution of comments. 

\begin{table}
\centering
\def\arraystretch{1.2}
\begin{tabular}{>{\hspace{0pt}}m{0.132\linewidth}>{\hspace{0pt}}m{0.163\linewidth}>{\hspace{0pt}}m{0.163\linewidth}>{\hspace{0pt}}m{0.002\linewidth}>{\hspace{0pt}}m{0.163\linewidth}>{\hspace{0pt}}m{0.19\linewidth}}
      & \multicolumn{2}{>{\centering\hspace{0pt}}m{0.366\linewidth}}{\textbf{Volume of Posting Activity}\par{}H4} &  & \multicolumn{2}{>{\centering\arraybackslash\hspace{0pt}}m{0.38\linewidth}}{\textbf{Distribution of Commenting Activity}\par{}H5}  \\ 
\cline{2-3}\cline{5-6}
      & \textbf{Model 8}\par{}vertex count\par{}(log) & \textbf{Model 9}\par{}edge count\par{}(log)&  & \textbf{Model 10}\par{}degree, avg.\par{} & \textbf{Model 11}\par{}degree, gini\par{}~     \\
      \cline{2-3}\cline{5-6}
      & $\beta$ (SE)    & $\beta$ (SE) &  & $\beta$ (SE)   & $\beta$ (SE)    \\ 
\hhline{~==~==}
\textbf{comm.~val.}     & 0.061 (0.044)   & .003~ (0.036)&  & 0.000 (0.001)  & 0.003(0.002)    \\
\textbf{content~restr.} & -0.001 (0.030)  & 0.015 (0.025)&  & -0.001 (0.001) & \textbf{-0.003 (0.001)}      \\
\textbf{struct.~reg.}   & \textbf{0.179* (0.069)}  & -0.027 (0.056)        &  & -0.001 (0.001) & -0.004(0.003)   \\ 
\hline
\textit{* p $<$ 0.05}   &        &     &  &       &       
\end{tabular}
\caption{Fixed-effect regression analysis results for submission networks. Independent variables are rule types (community values, content restrictions, and structural regulation) while Dependent variables are the respective network metrics.}
\label{table:subm-reg-analysis}
\end{table}

\section{Discussion}

The contribution of our work is two-fold. First, we contributed to further online community research by presenting a 15-code behavior-based rule classification of Reddit and three over-arching themes that cover Reddit's community rules. Second, we presented an analysis regarding how changes in these rule themes impact community discussion structure. Specifically, we found significant effects of:
\begin{itemize}
    \item Increase of Community Values leads to a notable increase of user interaction by 79\%, as does an increase of Content Restriction (57\%). 
    \item Increase of Structural Regulation leads to a 1\% more even distribution of user interaction throughout the community.
    \item Increase of Structural Regulation led to a ~5\% increase in clustering.
    \item Increase of Structural Regulation led to significantly greater posting activity (~18\%) in general.
    \item Increase of Content Restriction led to a small 0.3\% more even distribution of comments among posts.
\end{itemize}

Our study has several design implications as discussed below, including future work to better help creators and moderators regulate their online communities in effective ways. We note that, while this study is limited to using a subset of communities of Reddit and may not be generalizable to all other social network platforms, our work opens up opportunities for future research in this direction on other platforms with diverse, multi-faceted, and community-led moderation such as Wikipedia and Facebook groups \cite{Butler2008-aa, Seering2019-kx}.

\subsection{Relationships Between Community Rules and Member Behaviors}
We first discuss the implications of our study’s network analysis results.

A key finding of our work is that Structural Regulation rules are associated with both greater clustering and greater posting activity. Structural rules are often aimed towards helping members find topics of interest easier; these rules commonly direct posts to designated threads or even limit certain types of posts to particular times and days in the community. We found that although these rules indeed seem to raise posting frequency, they also cause greater subgroup formation within the community. For many community creators and moderators, enabling cliques to form may counter some community goals and lead away from a macro group-like culture in which members interact with the rest of the community equally. On the other hand, communities who wish to avoid clustering effects in the community may find benefit from avoiding Structural-related rules and/or instead implementing Community Values or Content Restrictions rules, which predict higher user interaction levels without any increase in clustering. 

We also discuss the emergence of sub- or even sub-sub-communities (such as clustering even within already “sub”reddits). Prior work has shown that smaller communities tend towards group-level dynamics, and confirmed our hypothesis that Structural-related rules lead to more even distribution of user interaction (along with higher clustering) [31]. Smaller communities often emerge as offshoots of larger communities due to users seeking more specific content. As a result, smaller communities may emerge from demand for a specific topic of interest to many people, and this narrowing of the space leads to a higher level of investment from members [31]. Smaller communities are more productive, supportive, and positive spaces. We note that, interestingly, we see this emerge as a result of Structural rules rather than Community Value-oriented rules (which had no significant relationship to distribution of user interaction).

It is worth highlighting a possible relationship between our definitions of Community Values and Content Restrictions. Specifically, the ideas of group interactions and topics of discussion may be closely related given prior work. For example, Sassenberg found that people who engage in more narrow topics (i.e. more restricted content) report greater group identity as a whole \cite{Sassenberg2002-qg}; however, groups that discuss a wide variety of topics reported other members as more personally likable, perhaps due to finding common bonds through sharing interactions and interests in a wide scope of people and topics. Our results somewhat corroborate Sassenberg’s findings that more restrictive topics and greater group identity are related, as we see user interactions increase with an increase in Content Restrictions (as well as Community Values). However, as opposed to Sassenberg’s findings that stronger group-level interactions and less personal-level interactions rise from a more narrow topic selection, our results do not find support for any differences in how interactions are distributed in the community given a rise in Content Restrictions. 

Surprisingly, we found that Content Restrictions did not show any negative relationships with user interaction or posting metrics. Interestingly, rules within this category are often the most strict and restrictive, such as specifically disallowing certain types of content and media or enforcing verification of information sources. It is possible that these restrictions do indeed demotivate users from participating in the community, but also increase user interaction from preventing bad behaviors between members. However, there is also evidence that content “restrictions”, which could also be interpreted as clarifying and defining a community’s allowed topics and values, do not actually deter contribution to a community but, in fact, encourage it. Kraut and Resnick found that people are only able to contribute content to a community once they are aware of its needs [43]. As Kraut and Resnick also note, making needed contributions easily visible (such as, in the case of Reddit, community rules explicitly present in a sidebar) increases likelihood that the community actually contributes. Additionally, we note that one of the most popular rules within the theme Content Restrictions is requiring on-topic content. Past work by Arguello et al. has shown that posting on-topic and specific questions increase reciprocated action like comments and replies [4]. However, we also note that other past work such as by Ren et al. has suggested that off-topic discussions can also provide more opportunity for relationships in a community; this is difficult to identify in our case given that peer relationships are only defined by commenting interactions given that Reddit does not have a “friend” or “follow” feature and users are largely anonymous. 

Overall, we find that our results show that, regardless of the rule themes, none of these rules showed a relationship with decreased user interaction. Increasing the number of rules did not result in any significant decrease in user interaction or posting activity. In other words, these promising results may show that “constraining” user behaviors on Reddit through more community-generated rules do not deter member participation in the community or form unnecessary barriers to users’ interactions.

\subsection{Proactive Moderation}
Past approaches to moderation in online communities have often focused on mitigating unwanted behaviors after they have occurred in the community. An important implication and future direction for our work is instead approaching online community moderation from a proactive and self-regulatory standpoint, and reducing the creation of unwanted online behavior through community regulation rather than responding to it. While reactive forms of moderation mitigate harm after it has already occurred, alienates community members, and disrupts the overall community environment \cite{Schluger2022-il}, proactive approaches to moderation prevent violating content before it ever occurs and are better-suited to promoting a healthy, civil community environment. Our work studying the effects of behavior-governing rules contributes to this largely unexplored but promising design space of proactive moderation. 

An explicitly proactive approach to moderation is overall less researched in HCI moderation. However, the concept has been explored by Seering, who posed questions on the potential for community-led moderation (such as that of Reddit) to encourage members to behave well in the first place \cite{Seering2020-gp}. As Seering notes, work on interface design and even “empathy nudges” have shown some success in promoting a more positive community environment. Schluger et al. found that moderators on Wikipedia pages already engage in proactive moderation behaviors, such as anticipating derailment during a conversation and intervening, and there is promise to automated processes for augmenting moderators’ workflow. However, there are many forms of proactive moderation used currently that are not scalable or practical in the long-term. For example, many online community moderators employ “pre-screening” on comments and posts; however, this is highly labor intensive and not sustainable for large communities \cite{Schluger2022-il}. As a result, proactive approaches to moderation that can encourage prosocial behaviors at scale are the most relevant to larger communities and platforms. As our study has shown, explicit community rules is one primary way to achieve this goal. We note that our work on behavior-governing rules is not only important for the direct influence on community members’ behaviors, but also to guide moderators’ roles as a “referee”; rules are important for reference by moderators when they do have to actively intervene in order to remind members of community values \cite{Schluger2022-il}. 

Additionally, although our work saw a large range of different rules across subreddits, we found that there exists common themes among the nearly 800 different community rules we explored as seen through our rule codebook's high-level rule themes and axial-codes. Our findings reveal that, despite hundreds of thousands of communities ranging widely in topic and separate moderator teams, Reddit's communities all share similar principles to one another in rule themes and focuses. One of the strengths of community-led moderation, such as on Reddit, is its ability to be context-sensitive and flexible to the local community, but this is in exchange for an immense amount of effort from individual members and moderators \cite{Seering2020-gp}. However, our work may show important implications for automated moderation tools even within these member-driven platforms, which can help reduce efforts of individuals. Through our work's findings of key principles in community rules that occur platform-wide even within the complex and community-driven ecosystems such as Reddit's, we hope that community designers and moderators may be able to further understand the potential effects of these rules on community interactions to help the development of moderation tools.

\subsection{Limitations}
Our work has the following limitations. First, our codebook was formed out of the top 100 subreddits and our network analysis included the top 2000 subreddits, among the over 100,000 active communities on Reddit. Although we ensured that our qualitative review of communities’ rules reached data saturation, our findings may not necessarily be representative of the inevitably complex and rich membership dynamics and rules of Reddit’s expansive platform. Second, our network analysis that involved scraping subreddits’ submissions and comments suffered from a small amount of missing data. To ensure that this did not impact our study’s findings, we manually sampled a random ten posts from a random 100 subreddits and found that less than 1\% of posts were missing. We did not observe significant trends in the missing posts themselves, such as missing more popular posts or concentrated from any particular subreddits. Given our findings, we do not expect this would have significantly changed our study’s results.

\section{Conclusion}
Our work investigated the relationship between communities’ rules and their network structure of discussion using Reddit, a popular social news sharing and discussion platform. We provided classification of behavior-based rules on Reddit as well as explored how a community’s implementation of certain types of rules affect users’ discussion structures. Our results indicate that this relationship has impactful effects for discourse online, and provide important implications in helping moderators and creators of communities to establish community rules that promote community goals. 

\bibliographystyle{ACM-Reference-Format}
\bibliography{ref}

\newpage
\appendix

\section{Subreddits used in thematic analysis for rule codebook}

\begin{figure}[h]
\includegraphics[scale=0.5]{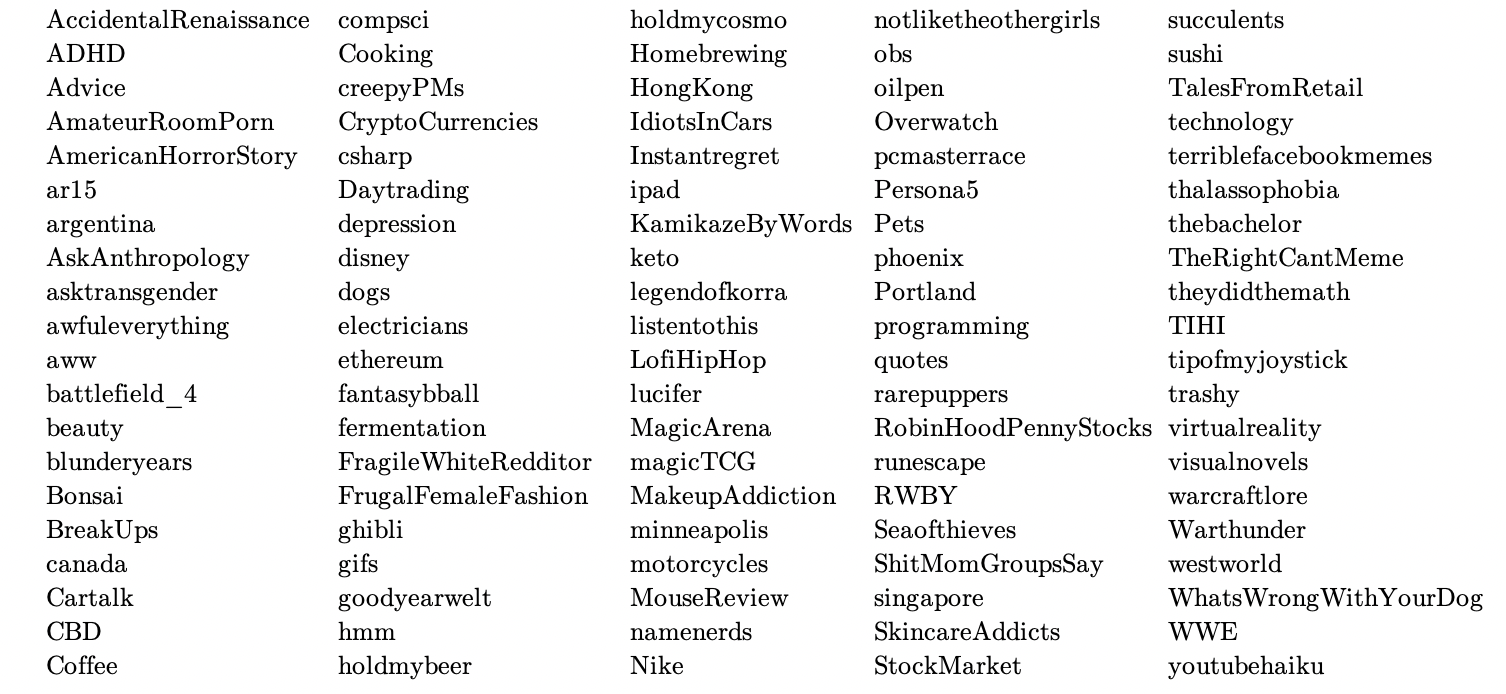}
\end{figure}
\section{Effects of changing the number of rules}

We found that there was a significant increase in clustering with more rules, as well as significantly greater user interaction with more than half an interaction increase with a single increase in the number of rules. Additionally, we found that power distribution between users becomes more uneven. We did not find any significant effects for the submission network. 

\begin{table}[h!]
\centering
\begin{tabular}{>{\hspace{0pt}}m{0.11\linewidth}>{\hspace{0pt}}m{0.08\linewidth}>{\hspace{0pt}}m{0.08\linewidth}>{\hspace{0pt}}m{0.08\linewidth}>{\hspace{0pt}}m{0.08\linewidth}>{\hspace{0pt}}m{0.1\linewidth}>{\hspace{0pt}}m{0.10\linewidth}>{\hspace{0pt}}m{0.1\linewidth}} 
\hline
   & \textbf{Model 1}\par{}vertex \par{}count\par{} & \textbf{Model 2}\par{}edge~\par{}count\par{} & \textbf{Model 3}\par{}density\par\null\par{} & \textbf{Model 4}\par{}clustering\par\null\par{} & \textbf{Model 5}\par{}degree,\par{}avg\par{} & \textbf{Model 6}\par{}degree, \par{}gini.\par{} & \textbf{Model 7}\par{}betw. centr.\par{}gini\par{}  \\
   & $\beta$~(SE)       & $\beta$ (SE)  & $\beta$ (SE)  & $\beta$ (SE)     & $\beta$ (SE)  & $\beta$ (SE)     & $\beta$ (SE)\\ 
\hhline{~=======}
\textbf{Number}\par{}\textbf{of rules} & 1821 \par{}(1071)      & 16846\par{}(16160)& 0.000 \par{}(0.000)        & 0.0053* \par{}(0.001)& 0.6229* \par{}(0.189)      & -.00076 \par{}(0.001)& .0004439* \par{}(0.000)  \\ 
\hline
* p $<$ 0.05   &      & & &    & &    &       
\end{tabular}
\end{table}

\begin{table}[h]
\centering

\begin{tabular}{>{\hspace{0pt}}m{0.11\linewidth}>{\hspace{0pt}}m{0.18\linewidth}>{\hspace{0pt}}m{0.18\linewidth}>{\hspace{0pt}}m{0.11\linewidth}>{\hspace{0pt}}m{0.11\linewidth}} 
\hline
   & \textbf{Model 8}\par{}\textbf{vertex count (log)} & \textbf{Model 9}\par{}\textbf{edge count (log)} & \textbf{Model 10}\par{}\textbf{degree,~avg.} & \textbf{Model 11}\par{}\textbf{degree,~gini.}  \\
   & $\beta$ (SE)       & $\beta$ (SE) & $\beta$ (SE)& $\beta$ (SE) \\ 
\hhline{~====}
\textbf{Number}\par{}\textbf{of rules} & 0.039 (0.020) & 0.016 (0.016)        & -0.001 (0.001)    & -0.000 (0.000)      \\ 
\hline
* p  0.05   &      &    & &  
\end{tabular}
\caption{Results for fixed-effect regression for user interaction network (top) and submission network (bottom) using independent variable of number of rules.}
\end{table}
\end{document}